\documentclass[conference,compsoc,a4paper]{IEEEtran}

\ifCLASSOPTIONcompsoc
  \usepackage[nocompress]{cite}
\else
  \usepackage{cite}
\fi

\usepackage{amssymb}
\usepackage{enumitem}
\usepackage{tcolorbox}
\usepackage[all]{nowidow}
\usepackage{siunitx}
\usepackage[linesnumbered,ruled,vlined]{algorithm2e}
\usepackage{multirow}
\usepackage{amsmath}
\usepackage{amssymb}
\usepackage{caption}
\usepackage{subcaption}
\usepackage{colortbl}
\usepackage{diagbox, eqparbox, hhline}
\usepackage[breaklinks]{hyperref}
\usepackage{booktabs}
\usepackage[T1]{fontenc}

\newcommand{\refappendix}[1]{\hyperref[#1]{Appendix~\ref*{#1}}}

\newcommand{\sandbox}{\texttt{The Privacy Sandbox}}

\newcommand{\topics}{\texttt{Topics}}
\newcommand{\browsingTopics}{\textit{<browsingTopics()>}}
\newcommand{\unknownTopic}{\textit{Unknown}}

\newcommand{\eg}{e.g.,}
\newcommand{\ie}{i.e.,}
\newcommand{\etc}{etc.}

\newcommand{\shortsection}[2][.]{\vspace{1mm}\noindent\textbf{#2#1}}

\definecolor{Gray}{gray}{0.9}

\newif{\ifanonymous}
\anonymousfalse{} % hide or show authors

\hypersetup{colorlinks, linkcolor=blue}
\DeclareSIUnit[number-unit-product=]\percent{%   % remove added space when using %
    \char`\%}

\usepackage{fancyhdr}

\begin{document}
\title{A Public and Reproducible Assessment of the \topics{} API on Real Data}

\fancypagestyle{firststyle}
{
   \fancyhf{}
    \chead{\textit{Revisions: simulation \href{https://github.com/yohhaan/topics_api_analysis/issues/1}{bug} fixed and new Topics classifier (v5), see updated quantitative results in this latest version}}
   \renewcommand{\headrulewidth}{0pt} % removes horizontal header line
}

\ifanonymous
\author{\em Anonymous Authors}
\else
\author{\IEEEauthorblockN{Yohan Beugin}
\IEEEauthorblockA{University of Wisconsin-Madison\\
Madison, USA\\
ybeugin@cs.wisc.edu}
\and
\IEEEauthorblockN{Patrick McDaniel}
\IEEEauthorblockA{University of Wisconsin-Madison\\
Madison, USA\\
mcdaniel@cs.wisc.edu}}
\fi

\maketitle
\thispagestyle{firststyle}
\pagestyle{plain}
\begin{abstract}
%Area
The \topics{} API for the web is Google's privacy-enhancing alternative to replace third-party cookies. Results of prior work have led to an ongoing discussion between Google and research communities about the capability of \topics{} to trade off both utility and privacy.
%Problem
The central point of contention is largely around the realism of the datasets used in these analyses and their reproducibility; researchers using data collected on a small sample of users or generating synthetic datasets, while Google's results are inferred from a private dataset.
%Solution
In this paper, we complement prior research by performing a reproducible assessment of the latest version of the \topics{} API on the largest and publicly available dataset of real browsing histories.
% Methodology
First, we measure how unique and stable real users' interests are over time. Then, we evaluate if \topics{} can be used to fingerprint the users from these real browsing traces by adapting methodologies from prior privacy studies. Finally, we call on web actors to perform and enable reproducible evaluations by releasing anonymized distributions.
%Results
We find that for the \num{1207} real users in this dataset, the probability of being re-identified across websites is of \SI{2}{\%}, \SI{3}{\%}, and \SI{4}{\%} after \num{1}, \num{2}, and \num{3} observations of their topics by advertisers, respectively.
%Take aways
This paper shows on real data that \topics{} does not provide the same privacy guarantees to all users and that the information leakage worsens over time, further highlighting the need for public and reproducible evaluations of the claims made by new web proposals.
\end{abstract}

% \begin{IEEEkeywords}
% Topics API, Privacy Sandbox, Re-identification, Fingerprinting, Web Behaviors
% \end{IEEEkeywords}

% \blfootnote{}

\section{Introduction}\label{introduction}

The \topics{} API from \sandbox{} is being developed by Google to deprecate third-party cookies on Chrome with the promise of both providing utility to advertisers and privacy to users~\cite{google_topics_2022}. \topics{} works by having the web browser classify the websites visited by users into categories of interest. Advertisers who are embedded on websites can observe some of the recent top users' topics and use that information to perform an ad auction~\cite{google_github_2022,google_topics_2023}. Prior analyses~\cite{thomson_privacy_2023,jha_robustness_2023,beugin_interest-disclosing_2024} have pointed at different limitations of the proposal. For instance, in a worst-case analysis quantifying the exact utility and privacy goals stated on the \topics{} proposal by Google, we demonstrated that users with stable interests have a higher risk of being re-identified across website visits. Indeed, over \SI{60}{\%} of the \num{250}k stable users that we simulated are not guaranteed strictly more than 10-anonymity after 10 observations~\cite{beugin_interest-disclosing_2024}.

These results have led to an ongoing discussion between Google and research communities about the capability of the \topics{} API to deliver on both its utility and privacy objectives. The major point of contention between the analyses carried out by Google and researchers is the access asymmetry to real browsing data as well as the resulting reproducibility of the evaluations. Indeed, while researchers have either collected browsing data on a small sample of \num{268} users~\cite{jha_robustness_2023} or synthetically generated large traces~\cite{beugin_interest-disclosing_2024}, Google performed their evaluations on a private dataset~\cite{epasto_measures_2022,carey_measuring_2023} and only reported aggregate results~\cite{thomson_privacy_2023}, making it impossible to reproduce Google's evaluation.

In this paper, we evaluate the latest version of the \topics{} API on the largest publicly available dataset of real browsing histories that we could find~\cite{kulshrestha_web_2020,kulshrestha_web_paper_2020}; complementing prior work and proposing an alternative to having to trust Google's non-reproducible assertions. We adapt prior methodologies to measure the fingerprinting potential of the \topics{} API on this publicly accessible dataset. Finally, we discuss future research avenues and call on web actors to release anonymized distributions to enable further reproducible analyses.

First, we measure on an anonymized dataset of over a month of real browsing histories how stable and unique users' online behaviors and interests are over time. This is to compare with a stability assumption assumed in prior work~\cite{beugin_interest-disclosing_2024}. Then, we adapt prior privacy analyses of the \topics{} API, but on the latest version of the proposal\footnote{Commit sha \href{https://github.com/patcg-individual-drafts/topics/tree/2df537b95e2e1377825d29121d560b0042387feb}{2df537b} as of July 8, 2024} as a new topics taxonomy, a new machine learning classifier, and other modifications of the topics computation have been released since. Finally, we simulate how the \topics{} API would work for these users and evaluate if advertisers can fingerprint and re-identify users through their topics.

We find that each week at least \SI{93}{\%} of the users in this dataset have a unique top \num{5} topics profile. Measuring how stable these top profiles are over time, we obtain that at least \SI{47}{\%} of users have \num{3} or more topics in common in their top \num{5} topics profile from one week to the next one, while only less than \SI{6}{\%} of users have none. The combination of unique and stable profiles opens the possibility to use the \topics{} API to fingerprint users. We observe that in practice for the \num{1207} real users from the dataset, their probability of being re-identified across \num{2} websites is of \SI{2}{\%}, \SI{3}{\%}, and \SI{4}{\%} after \num{1}, \num{2}, and \num{3} observations of their topics by third parties, respectively. This paper highlights the importance of public and reproducible evaluations of any claim made by new web proposals and to identify the potential limitations of these techniques during their design rather than after their deployment.

We make the following contributions:
\begin{itemize}
    \item The first privacy analysis of the newest version of the \topics{} API.
    \item The use of the largest publicly available dataset of real browsing histories to evaluate the \topics{} API.
    \item A reproducible methodology confirming on real data that users can be fingerprinted by their topics.
\end{itemize}

\section{Background \& Related Work}\label{background}
\label{relatedwork}

\subsection{\topics{} API for the Web}

\shortsection{Description} The \topics{} API for the web aims at deprecating third-party cookies while still enabling interest-based advertising~\cite{google_topics_2022,dutton_topics_2022,google_github_2022}. Google proposes that without third-party cookies, web browsers be the ones to classify users' online behaviors into topics of interests and return the top visited topics to advertisers embedded on the websites being browsed. As of February 2024, the \topics{} API for the web is being rolled out to every Chrome users~\cite{google_preparing_2023, google_shipping_2023}, while other browsers such as Apple Safari~\cite{annevk_topics_2022}, Mozilla Firefox~\cite{thomson_request_2022,thomson_privacy_2023}, or Brave~\cite{snyder_googles_2022} have declined to implement \topics{} in their product citing several privacy concerns with the proposal.

\shortsection{Privacy Claims} Google claims that the \topics{} API for the web makes it \textit{``difficult to reidentify significant numbers of users across sites using just the API''} and that \textit{``the topics revealed [are] less personally sensitive about a user than [...] today's tracking methods''}~\cite{google_github_2022}.

\begin{figure}
    \centering{}
    \includegraphics[width=.95\linewidth]{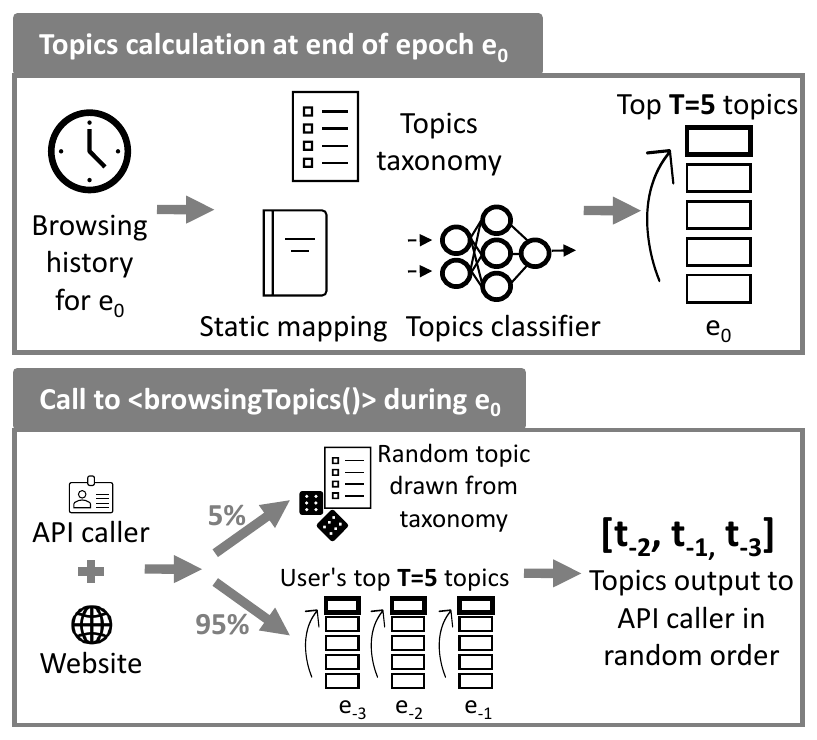}
    \caption{Overview of the \topics{} API for the web~\cite{beugin_interest-disclosing_2024}}
    \label{fig:topics}
\end{figure}

\shortsection{Technical Details} Next, we summarize how the \topics{} API is implemented; we adopt the notations (see \refappendix{notations}) introduced in our prior work~\cite{beugin_interest-disclosing_2024}.
\autoref{fig:topics} shows how at the end of an epoch $e_0$, the selection of the top $T=\num{5}$ most visited topics for each user is performed; the websites visited by users are classified locally by their web browsers into topics from \textbf{the taxonomy} using the static mapping and the machine learning model released by Google. \textbf{The static mapping} is a list annotated by Google of domains to their corresponding topics, \textbf{the \topics{} classifier} checks first if the domain is present in that list before invoking the ML model for classification. Then, when an advertiser embedded on a website calls the \topics{} API, an array of maximum $\tau=\num{3}$ topics and randomly shuffled is returned; for each of the last $\tau=\num{3}$ epochs, either with probability $p=\num{0.05}$, a random topic is drawn uniformly from the taxonomy, or with the opposite probability $q=1-p=\num{0.95}$, a real one is picked from the corresponding top $T=\num{5}$ most visited topics. \textbf{The witness requirement} in \topics{} enforces that a real topic is returned to API callers only if they were embedded on a website of the same topic that was also visited by the user during the past $\tau=\num{3}$ epochs.

\subsection{Analyses of the \topics{} API}

\shortsection{Google's Analyses} Google released two privacy analyses of the fingerprinting risk of the \topics{} API; a white paper computing the aggregate information leakage per API call and for two consecutive calls~\cite{epasto_measures_2022} and a second work using a theoretical framework to measure re-identification risk~\cite{carey_measuring_2023}. However, the empirical measurements in both works have been performed on a private dataset, preventing the verification of the claims being made. Similarly, only aggregate and final results are reported, the lack of details across the distribution of users can hide the privacy risks of \topics{} for specific users as already pointed out by Thomson~\cite{thomson_privacy_2023}. Additionally, Google's second analysis assumes and infers from an aggregate statistic that \textit{``for every user, samples of top sets [of topics] are independent across time''}~\cite{carey_measuring_2023}, while prior web measurement studies have found that users' interests exhibit some stability over time~\cite{greenberg_computer_1993,tauscher_how_1997,montgomery_identifying_2001,kumar_characterization_2010,goel_who_2012,tyler_large_2015,muller_understanding_2015}. In this paper, we discuss these assumptions and methodologies, argue the need for reproducible evaluations of the \topics{} API, and perform one on real data.

\shortsection{Independent Analyses} One of the first analysis of the \topics{} proposal was carried out by Thomson, an engineer at Mozilla, who outlined potential limitations and risks associated with implementing the API~\cite{thomson_privacy_2023}. Then, Jha et al., iterated on this initial analysis, collected browsing data on a sample of \num{268} users, and evaluated users' risk of being re-identified across websites through their topics of interest~\cite{jha_robustness_2023}. Concurrently, we performed a systematic evaluation of the \topics{} API for the web against Google's own stated goals, and uncovered how difficult it can be for the \topics{} API to deliver utility while guaranteeing privacy to all users~\cite{beugin_interest-disclosing_2024}. In our worst-case analysis of the \topics{} API, we showed how advertisers could identify at least \SI{25}{\%} of the noisy topics added by the API to its results and how users with stable interests are at a higher risk to be fingerprinted across websites through their topics. We found that over \SI{60}{\%} of the \num{250}k stable users we simulate are not guaranteed strictly more than \num{10}-anonymity for \num{10} observations of their topics by advertisers. In addition to finding these limitations to Google's privacy objectives, we also measured the accuracy of the \topics{} classification, informing on the level of utility retained by advertisers with this new API, and demonstrated how third parties can abuse the system by causing misclassifications~\cite{beugin_interest-disclosing_2024}.

\begin{figure*}[!ht]
    \centering
    \begin{subfigure}[b]{0.32\linewidth}
        \centering
        \includegraphics[width=\textwidth]{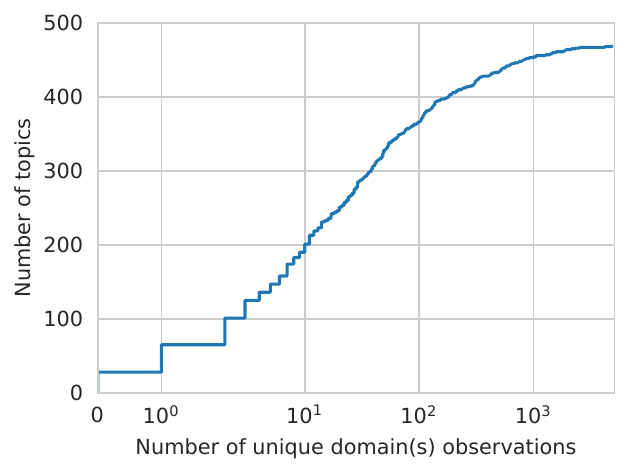}
        \caption{Static mapping}
        \label{fig:static_mapping_distribution}
    \end{subfigure}
    \hfill
    \begin{subfigure}[b]{0.32\linewidth}
        \centering
        \includegraphics[width=\textwidth]{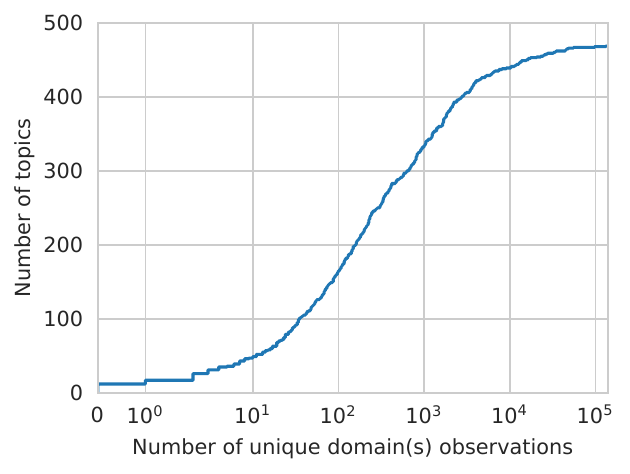}
        \caption{CrUX}
        \label{fig:crux_distribution}
    \end{subfigure}
    \hfill
    \begin{subfigure}[b]{0.32\linewidth}
        \centering
        \includegraphics[width=\textwidth]{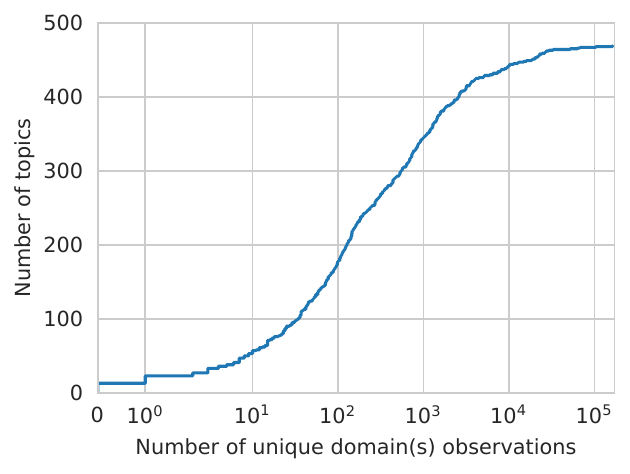}
        \caption{Tranco}
        \label{fig:tranco_distribution}
    \end{subfigure}
    \caption{Empirical cumulative distributions of the number of topics per number of unique domain(s) as observed on different lists of most visited domains.}
    \label{fig:distributions}
  \end{figure*}

\section{Methodology}\label{approach}

In the following section, we present the modifications that have been made to the \topics{} API since previous analyses were published and describe the dataset of real browsing histories on which we simulate \topics{}.

\subsection{New \topics{} API Version}
\label{new_version}

Since the last analyses of the \topics{} API~\cite{thomson_privacy_2023,jha_robustness_2023,beugin_interest-disclosing_2024}, Google made a few modifications to the implementation to improve its utility~\cite{google_topics_latest_2023}, we list them here, the versions used in this paper are also listed in \refappendix{notations}.

\shortsection{Modifications to Domains Classification}
First, a new taxonomy (version \textit{2}) of $\Omega=\num{469}$ topics has been released: \num{160} topics from the initial taxonomy have been removed and \num{280} new ones added. The model classifier (version \textit{5}) has been updated accordingly, and the static mapping that consists of a list of websites to topics manually annotated by Google has grown from about \num{10}k to now \num{50}k domains.

\shortsection{Modifications to Top Topics Selection}
In the newest version of the \topics{} API, the selection of the top $T=\num{5}$ topics for each epoch has been modified to not only take into account the frequency of each topic, but also their utility for ad relevance. Each topic from the taxonomy is now assigned a \textit{standard} or \textit{higher} utility label, and top topics are now selected by first ranking topics by utility and then frequency. This modification was made by Google after receiving feedback from advertisers that topics observed many times and shared by a lot of users (general topics like \textit{News} or \textit{Arts \& Entertainment}) were not so useful for advertising purposes~\cite{google_topics_latest_2023}. As a result, in the new version, topics of higher utility (less general, but also potentially more revealing privacy-wise) are considered before others.

\shortsection{Re-implementation and Topics Distribution} For this paper, we re-implement the latest version of the \topics{} API (commit sha \href{https://github.com/patcg-individual-drafts/topics/tree/78b13c3a3a0b14c82e35723402f5f742ce14422b}{78b13c3}) shipped in Google Chrome locally to be able to classify several thousands domains at once. This lets us obtain the prior distributions of topics observations on the most visited websites; \autoref{fig:distributions} shows such distributions on the following three lists of domains: the new static mapping used by Google composed of about 50k domains (\autoref{fig:static_mapping_distribution}), and the top 1M most visited domains as reported by CrUX~\cite{durumeric_cached_2023} (\autoref{fig:crux_distribution}) and Tranco~\cite{le_pochat_tranco_2019} (\autoref{fig:tranco_distribution}), respectively. While the \topics{} API has been updated since prior analyses, results about the asymmetry of the observations of topics from the new taxonomy on the most visited websites still hold~\cite{beugin_interest-disclosing_2024}; \num{28}, \num{12}, and \num{13} topics never appear at all from the taxonomy for the static mapping, CrUX, and Tranco, respectively, others only a few times, and very few on many domains from each list. Such prior distribution can be used by third party to distinguish if an observed topic is a noisy one (picked randomly from the taxonomy by the API in $p=\SI{5}{\%}$ of the cases) if it never appears or only seldomly on the top \num{1}M domains in practice (see \autoref{plausible_deniability_evaluation}).

\subsection{Dataset of Real Browsing Histories}

\shortsection{Initial Dataset} In this paper, we use data collected in October 2018 on \num{2148} users from Germany who have volunteered to share their desktop browsing histories in exchange for a financial compensation. The dataset has been shared online by researchers studying the habits~\cite{kulshrestha_web_paper_2020,kulshrestha_web_2020} and uniqueness~\cite{oliveira_browsing_2023} of web users activity. It was anonymized by only releasing the fully qualified domain name of the visited websites and not the complete URL of the pages browsed; thus avoiding to make public social media identifiers, personal profile pages, \etc{}~\cite{su_anonymizing_2017}. Overall, these \num{2148} users visited \num{9151243} URLs across \num{49918} unique domains or \num{67300} unique origins during \num{5} weeks.

\shortsection{Top Topics Profiles} We classify all the origins visited by the users during the \num{5} weeks from the initial dataset with the \topics{} classifier. This allows us to generate the top interests profiles of each user, exactly like the \topics{} API would do in practice if it were run on these browsing histories. We keep only users that have had a browsing history (and so topics assigned to them) for every week. Finally, we obtain a dataset composed of \num{1207} users who have visited \num{7746193} URLs across \num{43684} unique domains or \num{58370} unique origins in October 2018.

\section{Privacy Evaluation}\label{eval}

In this section, we perform a privacy analysis of the latest version of the \topics{} API on a dataset of real browsing histories. We start by outlining the privacy and utility trade-off of the \topics{} API.

\subsection{Privacy-Utility Trade-off}

\shortsection{Utility} The \topics{} API discloses to advertisers the interests that users had in the past epochs. Using these interests, third parties can infer which ads to display to users. Thus, the utility of the \topics{} API for advertisers is determined by how aligned the returned topics are with users' true interests. This is impacted by several factors such as the granularity of the taxonomy of topics, the accuracy of the classification of users' web behaviors by the API, the number of topics returned per API call and their nature (\SI{5}{\%} of the topics observed are expected to be noisy topics), the length of an epoch, as well as the number of websites on which the advertiser is embedded.

\shortsection{Privacy (k-anonymity)} Because the \topics{} API prevents advertisers from assigning unique identifiers to users through third-party cookies, Google claims that \topics{} is better for online users~\cite{google_github_2022}. \topics{} mainly relies on \textit{k-anonymity}; the idea is that advertisers would observe several users with the same topics of interest during an epoch, and so would not be able to distinguish and track them. In practice, \topics{} still discloses information about users to advertisers and one concern is that third parties use these signals for tracking. Users with stable and unique interests have a higher risk of being re-identified through their topics~\cite{thomson_privacy_2023,beugin_interest-disclosing_2024}. Indeed, although the API only discloses a maximum of \num{1} new topic per consecutive epoch after the initial call, advertisers can recover over time users' stable interests. These interest profiles collected over a longer period of time contain more information than just a single API call, and if they are unique enough they can be used to fingerprint users. Thus, in \autoref{stability}, we measure how stable and unique real users' interests are before evaluating their deanonymization risk in \autoref{reidentification_evaluation}.

\shortsection{Privacy (Plausible Deniability)} Google also claims that \topics{} does not disclose sensitive information about users because topics related to sensitive categories (\eg{} religion, politics, adult themes, \etc{}) have been removed from the taxonomy~\cite{google_github_2022}. This assumes that no topic in the current taxonomy can be associated with any sensitive subject which is not guaranteed; correlations between interests or within a specific context can be made~\cite{annevk_topics_2022,thomson_request_2022,thomson_privacy_2023}. Additionally, \topics{} aims to provide plausible deniability to users by adding random topics to \SI{5}{\%} of the results, so that users could blame on this randomness and ambiguity a topic observed by an advertiser rather than disclosing it was part of their browsing behavior~\cite{google_github_2022}. To improve the accuracy of their ads, advertisers will logically want to distinguish between noisy and true topics. We already established in \autoref{new_version} that the distribution of topics on the web was still not uniform at all. In \autoref{plausible_deniability_evaluation}, we re-evaluate on real data if this can be leveraged to identify noisy topics as in prior work~\cite{beugin_interest-disclosing_2024}.

\shortsection{Trade-off} As a result, there is fundamental trade-off between privacy and utility within the design of the \topics{} API; by disclosing users' interests directly to advertisers, the API also discloses some information about their browsing behaviors that can in some cases be used for tracking. Perhaps, the best illustration of this trade-off is the recent introduction of utility label along each topic of the taxonomy and the modification to how top topics are computed. Advertisers reported that during the origin trials of the \topics{} API, the results they were observing were too generic in scope and shared by many users (like \textit{News} or \textit{Arts \& Entertainment}). Because they were not so valuable to advertisers, Google modified the \topics{} API to output first topics with high utility rather than high frequency~\cite{google_topics_latest_2023}. If these changes aim to address utility issues faced by advertisers, they also impact users' privacy as the new topics now returned are potentially more unique and shared by less users.

\subsection{Analysis}

\shortsection{Research Questions} We ask the following questions:

\indent \textbf{(Q1)} How stable and unique are real users' profiles?

\indent \textbf{(Q2)} Can we flag noisy topics returned by real users?

\indent \textbf{(Q3)} Can we track across websites these real users?

\shortsection{Assumptions} To evaluate the privacy guarantees provided by the \topics{} API to these real users, we assume:

\begin{itemize}
    \item No witness requirement applied to API callers, \ie{} we consider that they can observe any topic. Note that this is aligned with the previous analyses of \topics{} including Google's~\cite{epasto_measures_2022,thomson_privacy_2023,jha_robustness_2023,beugin_interest-disclosing_2024,carey_measuring_2023}.
    \item Advertisers are observing and recording the topics of the same users across consecutive weeks. This lets them infer which topic likely corresponds to which week and so, we do not randomize the array returned by the API in our simulation.
\end{itemize}

\shortsection{Open Source Artifact} To enable replication of our results and further analyses, we release our code available at \url{https://github.com/yohhaan/topics_api_analysis}.

\subsection{Stability and Uniqueness of Topics Profiles}
\label{stability}

\begin{table*}
  \centering
  \caption{Number (and proportion) of users having 0 to 5 topics stable in their top 5 profiles for consecutive weeks.}
  \label{tab:stability_stats}
  \begin{tabular}{ccccccc}
  \toprule
   & 0 topic in common& Exactly 1 topic& Exactly 2 topics& Exactly 3 topics& Exactly 4 topics& Exactly 5 topics\\
  \midrule
  From week 1 to 2 &59 (4.9\%) &187 (15.5\%) &297 (24.6\%) &369 (30.6\%) &229 (19.0\%) &66 (5.5\%) \\
  From week 2 to 3 &70 (5.8\%) &189 (15.7\%) &318 (26.3\%) &345 (28.6\%) &226 (18.7\%) &59 (4.9\%) \\
  From week 3 to 4 &72 (6.0\%) &188 (15.6\%) &320 (26.5\%) &325 (26.9\%) &246 (20.4\%) &56 (4.6\%) \\
  From week 4 to 5 &70 (5.8\%) &240 (19.9\%) &324 (26.8\%) &318 (26.3\%) &211 (17.5\%) &44 (3.6\%) \\
  \bottomrule
\end{tabular}
\end{table*}

We seek to answer \textbf{(Q1)} by measuring how unique and stable the top topics profiles of the real users in our dataset are. Indeed, prior work showed that users with stable and unique interests in a population are at more risk of being re-identified by participating in the \topics{} API~\cite{thomson_privacy_2023,beugin_interest-disclosing_2024}.
\autoref{tab:stability_stats} reports how stable the top $T=\num{5}$ topics profiles of the real users in our dataset are across weeks. We can see from these results that very few users have top topics profiles totally unstable from week to week, but that rather users exhibit similar interests over time; less than \SI{6}{\%} of the users have \num{0} topic in common, while at least \SI{47}{\%} have \num{3} or more stable interests over time. \autoref{tab:dataset_stats} shows how many unique topics from the taxonomy are observed among all users each week; on this dataset of real browsing histories from 2018, only about \num{220} out of the \num{469} topics of the current taxonomy are present in the true topics of interest of users. \autoref{tab:dataset_stats} also reports for each week how many users have unique top $T=\num{5}$ topics profile; we observe that almost all of the \num{1207} users have unique topics profiles in this dataset. These results are aligned with prior ones in web measurement studies finding that web behaviors and interests of real users are unique~\cite{hutchison_analysis_2005,mcdaniel_system_2006,olejnik_why_2012,bird_replication_2020} and follow a recurrent system~\cite{greenberg_computer_1993,tauscher_how_1997,montgomery_identifying_2001,kumar_characterization_2010,goel_who_2012,tyler_large_2015,muller_understanding_2015}; users repeat actions leading to similar interests across weeks while also performing new activities. Thus, worst-case analyses assuming users' interests to be stable are important to understand the privacy limitations of the \topics{} API. Next, we adapt the privacy analysis of our prior work~\cite{beugin_interest-disclosing_2024} to measure these limitations on real data.

\begin{table}
  \centering
    \caption{Unique topics and top profiles across weeks.}
    \label{tab:dataset_stats}
    \begin{tabular}{cccccc}
      \toprule
      Week & 1 & 2 & 3 & 4 & 5\\
      \midrule
      Unique topics & 219 & 216 & 220 & 221& 226\\
      Unique profiles & 1127& 1132& 1142& 1143 & 1154\\
    \bottomrule
  \end{tabular}
\end{table}

\subsection{Identifying Noisy Topics}
\label{plausible_deniability_evaluation}

To address \textbf{(Q2)}, we hypothesize that some of the noise added $p=\SI{5}{\%}$ of the times by the API to the real topics observed by advertisers--in order to provide plausible deniability to users--can be removed by following a similar approach than in our prior work~\cite{beugin_interest-disclosing_2024}. This consists for advertisers in flagging topics that are never or seldomly observed on the web, but returned to them by the \topics{} API as noisy. Similarly, if advertisers determine that a topic is observed repeatedly for a user across different corresponding epochs, the topic is very likely to be a true one by the probabilistic nature of how the topics returned are picked.
To decide if a topic can be naturally observed on the web or not, we use the distribution of topics on the top \num{1}M most visited websites given by CrUX that we classified in \autoref{new_version}. We set a threshold of \num{10} domains on which each topic needs to appear to not be classified as noisy. This threshold corresponds to the one picked in our previous work~\cite{beugin_interest-disclosing_2024}, but we intend in future work to explore different thresholds and other approaches to try to identify noisy topics on this dataset of real browsing histories. We then simulate for each user and for each week \num{100} calls to the \topics{} API, flag topics we believe to be noisy, and compare the classification to the ground truth. We report in \autoref{tab:denoise_stats} the accuracy, precision, true positive rate (TPR) and false positive rate (FPR) of the classification of these \num{100} observations per user and per week (the positive and negative classes of our classifier are \textit{noisy} and \textit{real} topic, respectively). Note that the \topics{} API only outputs a noisy topic in about \SI{5}{\%} of the cases, as such this experiment presents a class imbalance, which is why we will mainly look at the \textit{precision} metric; the proportion of correctly flagged noisy topics among all instances flagged as noisy. Additionally, a topic observation requires knowing a user's top interests for the last $\tau =\num{3}$ weeks which explains why the table starts on week \num{3}. We find that with this approach, \SI{10}{\%} (precision) of the topics we flag are truly noisy which is better than \SI{0}{\%} if we were considering all topics to be genuine users' interests or the expected \SI{5}{\%} (remember the class imbalance) of a random guess. Thus, it is still possible in practice and with the latest version of \topics{} to refute plausible deniability to users for some topics. We leave to future work other classification heuristics.

\begin{table}
  \centering
    \caption{Classification results of the identification of the noisy topics returned by \topics{}.}
    \label{tab:denoise_stats}
    \begin{tabular}{ccccccccc}
      \toprule
      Week& Accuracy &Precision &TPR &FPR\\
      \midrule
      3 & 0.955 & 0.104 & 0.945 & 0.045\\
      4 & 0.955 & 0.103 & 0.936 & 0.045\\
      5 & 0.954 & 0.103 & 0.934 & 0.045\\
    \bottomrule
  \end{tabular}
\end{table}

\subsection{Re-identifying Users Across Sites}
\label{reidentification_evaluation}

\begin{figure}[!ht]
  \centering
  \includegraphics[width=\linewidth]{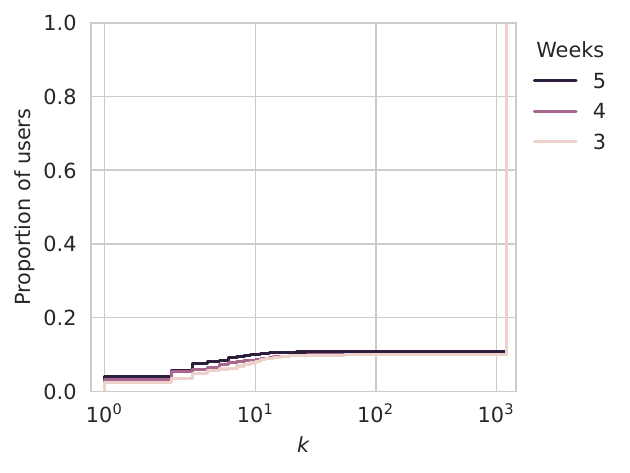}
  \caption{Cumulative distribution function of the proportion of users from the dataset that are provided $k$-anonymity. Note that $k=1$ means that users are re-identified across websites and so fingerprinted by the API.}
  \label{fig:5-week-reid}
\end{figure}

Finally, to respond to \textbf{(Q3)}, we run the following re-identification experiment~\cite{beugin_interest-disclosing_2024,carey_measuring_2023}. We consider that each user visits two different websites on which two different advertisers that are colluding are embedded. For each week that we simulate, these advertisers are observing the topics of each user from their respective vantage points. Their objective is to correlate a user seen by one advertiser on the first website with the correct user seen by the other advertiser on the second website. For each user, this matching is done by returning the user(s) that have the lowest Hamming distance between the topics observed by the second advertiser and the set of topics observed by the first advertiser~\cite{carey_measuring_2023}. We leave to future work the exploration of other matching strategies. This re-identification is performed every week with only the topics that have been observed until that week for each user. We present the results of this experiment in terms of level of $k$-anonymity that is provided to users by the API; \ie{}, $k=\num{1}$ means that advertisers have been able to correctly re-identify the user across their visits for that specific week, $1<k<n$ where $n$ is the number of users means that the user's visits could not be uniquely correlated to each other, but with the visits of $k-1$ other users, and $k=n$ means that the re-identification failed. Note that the experiment is also valid, and our results exactly the same, if we were to consider instead of two advertisers only one embedded on both websites and trying to correlate visits from both vantage points.
\autoref{fig:5-week-reid} shows for a simulation run the cumulative distribution function of the proportion of users that are satisfied $k$-anonymity (log scale) across the simulated weeks (recall that a topic observation for a user requires knowing their top interests for the last $\tau =\num{3}$ weeks, thus we start at week 3). Additionally, we report in \autoref{tab:reid_runs} the number of users that are correctly re-identified across observations for 10 simulation runs. We observe that the probability of being re-identified across visits for these users is of \SI{2}{\%}, \SI{3}{\%}, and \SI{4}{\%} after \num{1}, \num{2}, and \num{3} observations, respectively. For the \num{3} weeks simulated, we observe that about \SI{10}{\%} of the users are mapped to a group of size $k<n=\num{1207}$, which means that their cross-site visits can be correlated to each other with a higher probability than just a random guess. We conclude that some users from this real dataset are re-identified across websites through only the observations of their topics of interest in our experiment. Thus, the real users from our dataset can be fingerprinted through the \topics{} API. Moreover, as can be seen, the information leakage and so, privacy violation worsen over time as more users are re-identified.

\begin{table}
  \centering
    \caption{Number of users correctly re-identified after 1, 2, and 3 observations for several simulation runs.}
    \label{tab:reid_runs}
    \begin{tabular}{cccc}
      \toprule
       Simulation& 1 observation & 2 observations & 3 observations \\
       \midrule
       Run 1 &29 users &28 users &42 users  \\
       Run 2 &28 users &37 users &49 users \\
       Run 3 &29 users &47 users &63 users \\
       Run 4 &24 users &33 users &46 users \\
       Run 5 &30 users &39 users &52 users \\
       Run 6 &24 users &33 users &37 users \\
       Run 7 &31 users &38 users &49 users \\
       Run 8 &32 users &41 users &56 users \\
       Run 9 &29 users &27 users &48 users \\
       Run 10 &27 users &44 users &62 users \\ \hline
       \textbf{mean $\pm$ std} & \textbf{28.3 $\pm$ 2.7} & \textbf{36.7 $\pm$ 6.5} & \textbf{50.4 $\pm$ 8.2} \\
    \bottomrule
  \end{tabular}
\end{table}

\section{Discussion}\label{discussion}

\shortsection{Limitations} The dataset of real browsing histories used in this paper likely contains some biases; it was collected in Germany in October 2018 and so does not necessarily represent every individual browsing the web today. Additionally, its size remains somewhat limited in number of users or collected weeks. Nevertheless, the use of this dataset does inform us on the privacy guarantees that \topics{} would have delivered in practice to these specific real users if the API had been available at the time the data was collected. This is perhaps the most relevant in our endeavor to address the need to trust Google's assertions and results made on their private dataset. Indeed, this is, to the best of our knowledge, the largest publicly available dataset of real web histories on which a reproducible privacy evaluation of the \topics{} API has been performed.

\shortsection{Recommendations for Reproducible Evaluations} First, in order to evaluate the privacy and utility guarantees of a proposal, its goals should be clearly defined to be provable or quantifiable; in prior analysis, we had to rephrase the initial objectives stated by Google on the \topics{} API proposal as they were too vague~\cite{beugin_interest-disclosing_2024}. Second, one should avoid performing non-reproducible evaluations on a private dataset~\cite{epasto_measures_2022,carey_measuring_2023} as it limits the scope of these analyses to having to trust assertions being made without the possibility to verify them. If an evaluation on a private dataset needs to be performed, we would recommend publishing rigorous details about the experiment and the results rather than just final and aggregated metrics; indeed, Thomson already demonstrated the danger of reporting the average case rather than the full distribution on the privacy claims of the \topics{} API~\cite{thomson_request_2022,thomson_privacy_2023}. To solve this problem, we would like to call on web actors like Google--that have access to real browsing histories--to release the topics of interest of a representative population of users across several months, so that it can serve as a common starting point for future analyses. Note that such dataset not necessarily needs to include the browsing histories of users, but just the set of top topics of anonymized users. It could even be synthetically generated by drawing these topics profiles from a representative distribution of real users to address potential privacy concerns by releasing real data.

\shortsection{Future Work} Overall, rigorous and thorough analyses of these new web proposals (\sandbox{} for the web now encompasses more than 20 different mechanisms~\cite{google_privacy_2021}) are needed; several interesting questions remain unsolved, such as if these proposals can deliver on their guarantees to all users, if their specification and implementation are correct, if they can be manipulated or abused, if their results can be correlated with other fingerprinting signals, if developers and users do understand how these new APIs work, \etc{}
Regarding the \topics{} API, evaluating it without having access to Google's private dataset of browsing histories can be quite challenging; we refer readers to previous analyses of the \topics{} API to see how prior authors (partially) got around that obstacle~\cite{jha_robustness_2023,beugin_interest-disclosing_2024}. In this paper, we call on web actors to release anonymized or synthetic datasets to help with that point. In the meantime, we envision to perform a literature review and a taxonomy of web users measurement studies, pair these results with the few datasets of browsing histories publicly available, and generate synthetic distributions of users' topics of interest.

\section{Conclusion}\label{conclusion}

This paper complements previous analyses of the \topics{} API for the web by confirming on real data that the proposal does not provide the same privacy guarantees to all users. Specifically, we show how these real users exhibit stable and unique topics profiles and how they risk to be fingerprinted by their topics as the information leakage worsens over the time period that can be simulated with this dataset; the probability for these users to be re-identified across websites is of \SI{2}{\%}, \SI{3}{\%}, and \SI{4}{\%} after \num{1}, \num{2}, and \num{3} observations of their topics, respectively. 
We also highlight the need for reproducible analyses of the claims made by new web proposals before their deployment. This is important to avoid replicating the same mistakes already made in the past with the same technologies being deprecated today; the introduction of \eg{} web cookies later extensively abused for tracking purposes. As such, we call on web actors and industry to release representative and anonymized datasets to enable reproducible evaluations of the \topics{} API and other proposals from \sandbox{} over time.

\ifCLASSOPTIONcompsoc
  \section*{Acknowledgments}
\else
  \section*{Acknowledgment}
\fi
\ifanonymous
Anonymized for review.
\else
We sincerely thank the reviewers for their constructive comments and suggestions on this paper. We also thank Josh Karlin, lead engineer from Google on the \topics{} API for the help provided when pointing out a subtle \href{https://github.com/yohhaan/topics_api_analysis/issues/1}{bug} in our initial simulation for this paper; the code and quantitative results have been revised in this latest version.

\textbf{Funding acknowledgment:}
This material is based upon work supported by the National Science Foundation under Grant No. CNS-232088 and Grant No. CNS-2343611. Any opinions, findings, and conclusions or recommendations expressed in this material are those of the author(s) and do not necessarily reflect the views of the National Science Foundation.

\fi

\bibliographystyle{IEEEtran}
\bibliography{refs}

\appendices
\section{Notations}
\label{notations}

\begin{table}[!h]
    \centering
    \caption{Notations and symbols used in this paper.}
    \label{tab:freq}
    \begin{tabular}{ccc}
      \toprule
      Notation&Definition&Value\\
      \midrule
      Commit on \topics{} proposal & Commit sha of the \topics{} proposal
      studied in this paper
      &\textit{\href{https://github.com/patcg-individual-drafts/topics/tree/2df537b95e2e1377825d29121d560b0042387feb}{2df537b}}
      (July 8, 2024) \\
      Taxonomy version & Version of the taxonomy used in this paper & \num{2}\\
      Model version & Version of the studied \topics{} classifier in this paper&
      \num{5}\\
      Utility buckets version & Version of the studied \topics{} utility buckets in this paper&    1\\
      CrUX version & Version of the CrUX top-list used in this paper
      &\href{https://github.com/zakird/crux-top-lists/raw/main/data/global/202406.csv.gz}{202406}
      (June 2024) \\
      Tranco version & Version of the Tranco top-list used in this paper
      &\href{https://tranco-list.eu/download/G6KQK/1000000}{G6KQK} (August 3,
      2024) \\
      \browsingTopics{} & \topics{} API call &\textit{document.browsingTopics()}
      \\
    %   $A,B$ & Advertisers $A$ and $B$ & - \\
    %   $\mathcal{B}$ & Binomial distribution & - \\
    %   $e_i$ & Epoch $i$ & Size $e_{i+1}-e_i = 1~\text{week}$  \\
    %   $i,j$ & Generic math variables used for iterations & - \\
      $n$ & Number of users& -\\
      $p$ & Probability to pick a random (noisy) topic from taxonomy & \num{0.05}\\
      $q=1-p$ & Probability to pick a real topic from user's top $T$ topics &
      \num{0.95}\\
      $T$ & Number of top topics per epoch & \num{5} \\
      $\tau$ & Number of topics returned by \browsingTopics{} & \num{3} maximum \\
      $t_j$ & Topic $j$ & -\\
    %   $u_{i,A}$ & User of identity  $i$ observed by advertiser $A$& -\\
    %   $w_B$ & website on which advertiser $B$ is embedded & -\\
      $\Omega$ & Number of topics in taxonomy & \num{469} topics (+ \unknownTopic{}
      topic)\\
    \bottomrule
  \end{tabular}
\end{table}

\end{document}